\documentclass[journal]{IEEEtran}
\ifCLASSINFOpdf

\else

\fi
\usepackage{cite}
\usepackage{amsmath}
\usepackage{amssymb}
\usepackage{color}
\usepackage{soul}
\usepackage{graphicx}
\usepackage[ruled]{algorithm2e}
\usepackage{caption}
\usepackage{threeparttable}
\usepackage{amsthm}
\usepackage{theoremref}
\usepackage{color}
\makeatletter 
\g@addto@macro{\@algocf@init}{\SetKwInOut{Parameter}{Parameters}} 
\newcommand{\removelatexerror}{\let\@latex@error\@gobble}
\makeatother
\begin{document}
	
	\title{A novel online scheduling protocol for energy-efficient TWDM-OLT design}

	\author{Sourav~Dutta, Dibbendu~Roy, Chayan~Bhar
		and~Goutam~{Das}
		\thanks{Sourav Dutta is with the Department
			of Electronics and Electrical Communication Engineering, Indian institute of technology Kharagpur,  Kharagpur,
			India (e-mail: sourav.dutta.iitkgp@gmail.com).}
		\thanks{{D}ibbendu Roy, Chayan Bhar and Goutam Das are with G. S. Sanyal School of Telecommunication, Indian Institute of Technology Kharagpur, Kharagpur, India (e-mail: chayanbhar88@live.com, gdas@gssst.iitkgp.ernet.in).} 
		}

	\maketitle

	\begin{abstract}
		
		Design of energy-efficient access networks has emerged as an important area of research, since access networks consume $80-90\%$ of the overall Internet power consumption. TWDM-PON is envisaged to be one of the widely accepted future access technologies. TWDM-PON offers an additional opportunity to save energy at the OLT along with the existing energy-efficient ONU design. In this paper, we focus on the energy-efficient OLT design in a TWDM-PON.  While most of the conventional methods employ a minimization of the number of wavelengths, we propose a novel approach which aims at minimizing the number of voids created due to scheduling. In the process, for the first time, we present a low-complexity on-line scheduling algorithm for the upstream traffic considering delay constraints. Our extensive simulations demonstrate a significant improvement in energy efficiency of $\sim 25\%$ for high load at the OLT receivers. Furthermore, we provide an analytical upper-bound on the energy-efficiency of the OLT receivers and demonstrate that the proposed protocol achieves an energy efficiency very close to the bound with a maximum deviation  $\sim 2\%$ for $64$ ONUs.
	\end{abstract}
	
	\IEEEpeerreviewmaketitle
	\section{Introduction}
	
\begin{figure}[h]
		\centering
		\includegraphics[scale=1]{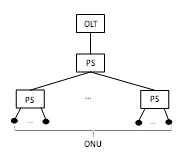}
		\caption{TWDM-PON architecture}
		\label{arc}
	\end{figure}

	The continuous increase in the number of Internet users has resulted in a subsequent increase in energy consumption. Therefore, obtaining an overall lower energy consumption figure has become the design objective of the present-day network engineers. The access network consumes a larger fraction ($80-90\%$) of the overall Internet power consumption \cite{kani}. Hence, energy-efficient access network design is drawing a lot of attention within the Information and Communication Technology (ICT) research community \cite{bhar2015designing,doze,dixit,Dixit:12,dixit2013towards,khotimsky2014unifying,twdm,online,twdm1,twdm2,diajocn,verfy,valcarenghi2015erratum}. The Passive Optical Network (PON) has emerged to be one of the widely accepted and deployed energy-efficient access solution. The access technology adopted in PON can be time (TDM-PON), wavelength (WDM-PON) or both time and wavelength multiplexed (TWDM-PON).  Among them, TDM-PON suffers from the absence of scalability whereas the opportunity of statistical multiplexing is absent in WDM-PON. To solve these drawbacks, Full Service Access Network (FSAN) group has opted for TWDM-PON as the future access solution  \cite{luo2012wavelength}. A PON consists of an Optical Line Terminal (OLT), multiple Optical Network Units (ONUs) with tunable transceivers and multiple stages of Remote Nodes (RNs) realized using passive power splitters (Fig. \ref{arc}).  Data from the OLT to ONUs (downstream) is broadcasted while data from ONUs to the OLT (upstream) is statistically multiplexed (centrally supervised by the OLT). The absence of collision domain in the downstream (DS) makes the upstream (US) traffic scheduling more challenging as compared to DS counter part. 	
	
	The OLT and ONUs, being the only active components, are solely responsible for energy consumption in TWDM-PON.  The protocols proposed for energy-efficient ONU design in TDM-PON \cite{bhar2015designing,doze,dixit,Dixit:12,dixit2013towards,khotimsky2014unifying} can also be extended for TWDM-PON \cite{twdm}. The unavailability of upstream data especially at low load creates a lot of idle periods between two consecutive US data transmission at the OLT receivers. These idle periods also known as voids provide an additional opportunity to save energy at the OLT as well. This motivates researchers towards designing energy-efficient OLT receivers in TWDM-PON.
	
	 A few Dynamic Wavelength and Bandwidth Allocation (DWBA) protocols for energy efficiency have already been proposed \cite{online,twdm1,twdm2,diajocn,verfy}. 	
	The basic methodology employed for utilizing those voids towards energy-efficiency at OLT receivers is same in all the existing proposals. The authors of \cite{online,twdm1,twdm2,diajocn,verfy} have focused on scheduling ONUs within fewer number of wavelengths termed as active wavelengths, and the rest of the wavelengths (inactive wavelengths) are utilized for energy-efficiency. This has directed the existing protocols towards minimizing the number of active wavelengths. Separate algorithms are employed for dynamic wavelength allocation (DWA) and dynamic bandwidth allocation (DBA) . DWA is performed only at some predetermined time instants (DWA boundaries), and is responsible for calculating the number of active wavelengths over which scheduling is performed. The calculation of the number of active wavelengths requires traffic prediction. DBA has the functionality of scheduling ONUs within the active wavelengths. The authors of \cite{online} and \cite{eft} employ Earliest Finish Time (EFT) and Latest Finish Time (LFT) respectively, for scheduling among these active wavelengths. However, in these protocols, delay constraint has not been considered which is a necessary requirement for delay sensitive traffic (like voice, video etc. \cite{del}). The authors of \cite{twdm1,twdm2} formulate an optimization problem for minimizing the number of active wavelengths while satisfying a delay constraint. A heuristic Dynamic Wavelength and Bandwidth Assignment (DWBA) has been proposed in \cite{diajocn}, in which the authors have analyzed the average delay and energy-efficiency figures. The obtained performance has been verified using simulation in \cite{verfy}.

	The conventional approach of minimization of the number of active wavelengths suffers from the under-utilization of the idle periods present at the OLT receivers since the voids created within the active wavelengths are not considered for energy-efficiency. Although, at the first glance it creates the impression that utilization of those voids for energy-efficiency can provide a minor improvement; a closer look at the protocol description reveals opportunity for significant improvement. The reason can be explained as follows: 
	
	The occupancy of the OLT receivers by the US data is defined as the utilization factor, $\rho$ which is the ratio of average arrival rate and average service rate. The reduction of the number of active wavelengths in the conventional approach results in a decrement of the available bandwidth (service rate) used for US data transmission. This leads to increase in the value of $\rho$ and hence, yields to creation of fewer (number of) or smaller(duration of) voids. When $\rho=1$ no voids are created. However, it is known that the average delay is very high even for Poisson traffic \cite{Data} if the value of $\rho$ is close to $1$. Therefore, the satisfaction of delay bound of the US data requires to restrict $\rho$ to values lower than $1$. In practice, the bursty nature of traffic imposes further restriction on $\rho$ to be much lower than $1$. Hence, the duration of voids created within the active wavelengths become high and their utilization can significantly improve the energy-efficiency. This motivates us to look for an alternate approach which utilizes the voids created due to scheduling for energy-efficiency.
	
	In this paper, we deviate from the conventional methodology of increasing the effective utilization factor by minimizing the number of active wavelengths. In our proposed methodology, all the wavelengths are kept on for scheduling (all wavelengths are active wavelengths), and hence, the utilization factor remains  unchanged (contrary to the conventional approach). This leads to creation of large amount (duration) of void(s) at the OLT receivers. These voids are utilized for energy-efficiency by switching off the OLT receivers within the created voids in our proposed approach. However, if the receivers of the OLT are switched off within the voids,  then the OLT requires some time to switch them on, which is known as sleep-to-wake–up time ($T_{sw}$). In this paper, we focus on the minimization of this loss due to $T_{sw}$ for further improvement of the energy efficiency. This is possible by minimizing the number of voids since the loss due to $T_{sw}$ is directly proportional to the number of voids if all voids are used for energy-efficiency (\ref{ssec:moti}). The only way to minimize the number of voids is by clubbing many small voids to form a bigger void (much larger as compared to $T_{sw}$) where the receivers of the OLT can be switched off for achieving energy-efficiency. Thus in our proposed approach, we minimize the number of voids by consolidating the US data transmissions of the ONUs. Since, the proposed methodology keeps the effective utilization factor to the lowest possible value by utilizing all the wavelengths for scheduling, it provides the highest opportunity of combining the US data  while maintaining the delay bound. This opportunity is absent in case of the conventional approach as the effective utilization factor is made very high by minimizing the number of active wavelengths.  The following advantages over the conventional methodology can further be achieved by employing the proposed approach: 
	\begin{itemize}
		\item 	The achieved energy-efficiency ($\eta$) by employing the number of active wavelength minimization schemes is upper limited by $\frac{W-1}{W}\times100\%$ ($W$- Number of wavelengths) since at least one wavelength requires to be always active ($\eta=50\%$ for $W=2$). However, the achieved energy-efficiency of our proposed algorithm is much higher than this upper bound (Section V (B)).
		\item The proposed approach eliminates the necessity of traffic prediction which is extremely essential in the conventional approach for calculating number of active wavelengths.
		\item	The proposed approach can react to the instantaneous   load change much faster than the conventional process as wavelengths are always active for scheduling in the proposed approach. Therefore, this leads to better satisfaction of QoS requirement.
	\end{itemize}

	 In this paper, we propose a novel online Dynamic Wavelength and Bandwidth Allocation (DWBA) scheme (EO-NoVM- Energy-efficient OLT design by employing number-of-voids minimization) considering a specified maximum delay bound. A significant energy benefit (maximum of $\sim 25\%$ at high load) has been demonstrated by employing EO-NoVM through extensive simulations (\ref{ssec:compb}). Further, we have analytically derived an upper bound of energy-efficiency that can be achieved from the OLT receivers. Through simulation, we demonstrate that the deviation of the achieved energy-efficiency from that analytical upper bound is restricted within $2\%$ for $N=64$ ($N$- number of ONUs).

	The rest of the paper is organized as follows. The proposed EO-NoVM algorithm is described in Section \ref{pro}. The complexity analysis of EO-NoVM is presented in Section \ref{comp}. In Section \ref{maxcal}, we calculate an upper bound for energy-efficiency as a function of the traffic load. In Section \ref{rd}, we present simulation results followed by brief discussions on them. Finally, concluding remarks are presented in Section \ref{conc}.
	\section{Proposed protocol}\label{pro}
	In this section, we propose an online scheduling scheme for the design of energy-efficient OLT (EO-NoVM) for TWDM-PON. A brief motivation behind the proposed EO-NoVM is provided in the first subsection. This is followed by a discussion of the proposed EO-NoVM algorithm in the next subsection. 
	\subsection{Motivation}\label{ssec:moti}
	Here, we present the intuition behind our proposed EO-NoVM scheme. Let us consider a very long observation duration $T_{obs}$ ($T_{obs} \rightarrow \infty$). We observe that two different methods exist for the improvement of energy-efficiency over the $T_{obs}$ duration if the receivers of the OLT are switched off within each void for energy-efficiency. 
	Firstly, the total duration of void over $T_{obs}$  due to the unavailability of the US traffic depends on the utilization factor $\rho$, defined as the ratio of average arrival rate and average departure rate. If $r_a^i$  is the average arrival rate of $i^{th}$ ONU ($ONU_i$) then the total average arrival rate of all $N$ ONUs is given by; $\sum_{i=0}^{N}r_a^i$. Further, if a single wavelength of the feeder fiber provides a bandwidth (data rate), $r_d$ then the overall data rate achieved from all $W$ wavelengths is given by; $Wr_d$. Hence, $\rho$ is given by; $\sum_{i=0}^{N}r_a^i/Wr_d$.  The total duration within $T_{obs}$ over which no US data is transmitted is given by; $(1-\rho) T_{obs}$. In a typical Muti-Point Control Protocol (MPCP) employed for DBA \cite{ieee2010ieee}, ONU sends a REPORT message after each US data transmission to inform the OLT about their current buffer states. The OLT allocates (schedules) a US transmission slot for an ONU based on its reported buffer state by sending a GATE message. Further, a guard duration ($t_g$) exists between two consecutive US data transmissions \cite{ieee2010ieee}. Let us assume that $M$ number of REPORT messages and guard durations are present in $T_{obs}$ duration. If $N_R$ and $T$ are the size of the REPORT message (in Bytes) and the byte duration respectively, then the average duration of voids that is wasted for the transmission of REPORT messages and guard durations is given by; $M(N_R T+T_g)$. Hence, the total duration of voids ($T_v^{agg}$)  created over the $T_{obs}$ duration is given by;  
	$$T_v^{agg}=(1-\rho) T_{obs}-M(N_R T+T_g )$$
	Since, all the terms of $T_v^{agg}$ except $M$ are constant for a fixed arrival rate of the ONUs, increment of $T_v^{agg}$ is only possible by decreasing $M$. The minimization of $M$ requires OLT to schedule the already reported ONUs to the latest time instant possible while maintaining the delay constraint.
	
	The other possibility for improving energy-efficiency is by creating the voids within scheduling process in such a way that the energy-efficiency approaches maximum. This is explained as follows. We have already mentioned that if the receivers of the OLT are switched off then $T_{sw}$ duration is wasted in each void. Since OLT itself is responsible for scheduling and the voids are only created by the scheduling process, OLT has the prior information about the exact time instants and the duration of all the voids. Therefore, the entire inactive period of the OLT receivers can be exploited for energy-efficiency.  Since each void contains exactly one $T_{sw}$ and all of them are utilized for energy-efficiency, the loss of energy-efficiency due to $T_{sw}$ is directly proportional to the number of voids. Number of voids can only be reduced by combining them to form larger voids. This can be realized by clubbing the US data transmissions of different ONUs.
	
	The above discussion reveals two possible directions towards the design of scheduling algorithm for energy-efficient OLT. 
	\begin{itemize}
		\item Schedule to the latest time instant possible while satisfying the delay constraint (to reduce M)
		\item 	Schedule in such a way that the clubbing of the US data is maximized (to minimize the loss due to $T_{sw}$)
	\end{itemize} 
	\begin{table}
		\caption{Definition of notation}
		\begin{center}
			\begin{tabular}{|c|p{7cm}|}
				\hline
				\multicolumn{1}{|c|}{\textbf{Notation}} & 
				\multicolumn{1}{c|}{\textbf{Description}}\\ \hline
				
				$N$ & Number of ONUs \\
				$W$ & Number of wavelengths\\
				$T$ & Byte duration\\
				$N_R$ & Size of the REPORT message (Bytes)\\
				$T_g$ & Guard duration\\
				$T_p$ & Gate message generation time\\
				$T_{tx}$ & Gate message transmission time\\
				$T_t^{i,j}$ & Tuning time between wavelength $i$ and $j$\\
				$T_{sw}$ & Sleep-to-wake-up time\\
				$T_{rtt}^k$ & round-trip time of $ONU_k$\\
				$D_{max}^k$ & Maximum allowable packet delay of $ONU_k$\\
				$D_q^k$ & 	Delay constrain of 〖$ONU_k$ after receiving the $q^{th}$ REPORT message by the OLT\\
				$G_k$ & Grant size of $ONU_k$\\
				$V$ & Set of all voids\\
				$LF$ & Set of latest scheduling horizon of all wavelength\\
				$v_i^s$ & Start-time of $i^{th}$ void\\
				$v_i^e$ & End-time of $i^{th}$ void\\
				$v_i^w$ & Wavelength of $i^{th}$ void\\
				$ lf_j$ & Latest scheduling horizon of wavelength $j$\\
				$T_{sch}^k$ & The time instance of generation of GATE message for $ONU_k$\\
				$W_{sch}^k$ & Scheduled wavelength of $ONU_k$
				\\ \hline 
			\end{tabular}
		\end{center}
	\end{table}
	
	Scheduling towards the latest time instant not only reduces $M$ but also creates possibility of reduction in the number of voids since each schedule, after receiving the REPORT message, has the potential to create exactly one void. However, scheduling at latest time may create an extra void which may be avoided by scheduling it little earlier. Further, the improvement of energy-efficiency by reducing $M$ is very small as compared to the improvement of energy-efficiency achieved by reducing the number of voids. This is due to the fact that  $T_{sw}$ ($\sim ms$ \cite{diajocn}) is much larger than $N_R T+T_g$ ($\sim \mu s$). Therefore, a greedy method is employed in EO-NoVM for minimization of the number of voids.  Hence, in EO-NoVM, the OLT tries to schedule an ONU ($ONU_k$) after receiving the REPORT from it (online) in such a way that the US transmission of $ONU_k$ either initiates immediately after the termination of some other ONUs (already scheduled) US transmission window or terminates immediately before the beginning of some other ONUs US transmission. If multiple such possibilities are available or if no such clubbing of US data is possible then DWBA is focused towards minimizing $M$ (among those possibilities) which is realized by scheduling at the latest time instant possible while satisfying the delay constraint.
	
\begin{figure}
		\centering
		\includegraphics[scale=0.38]{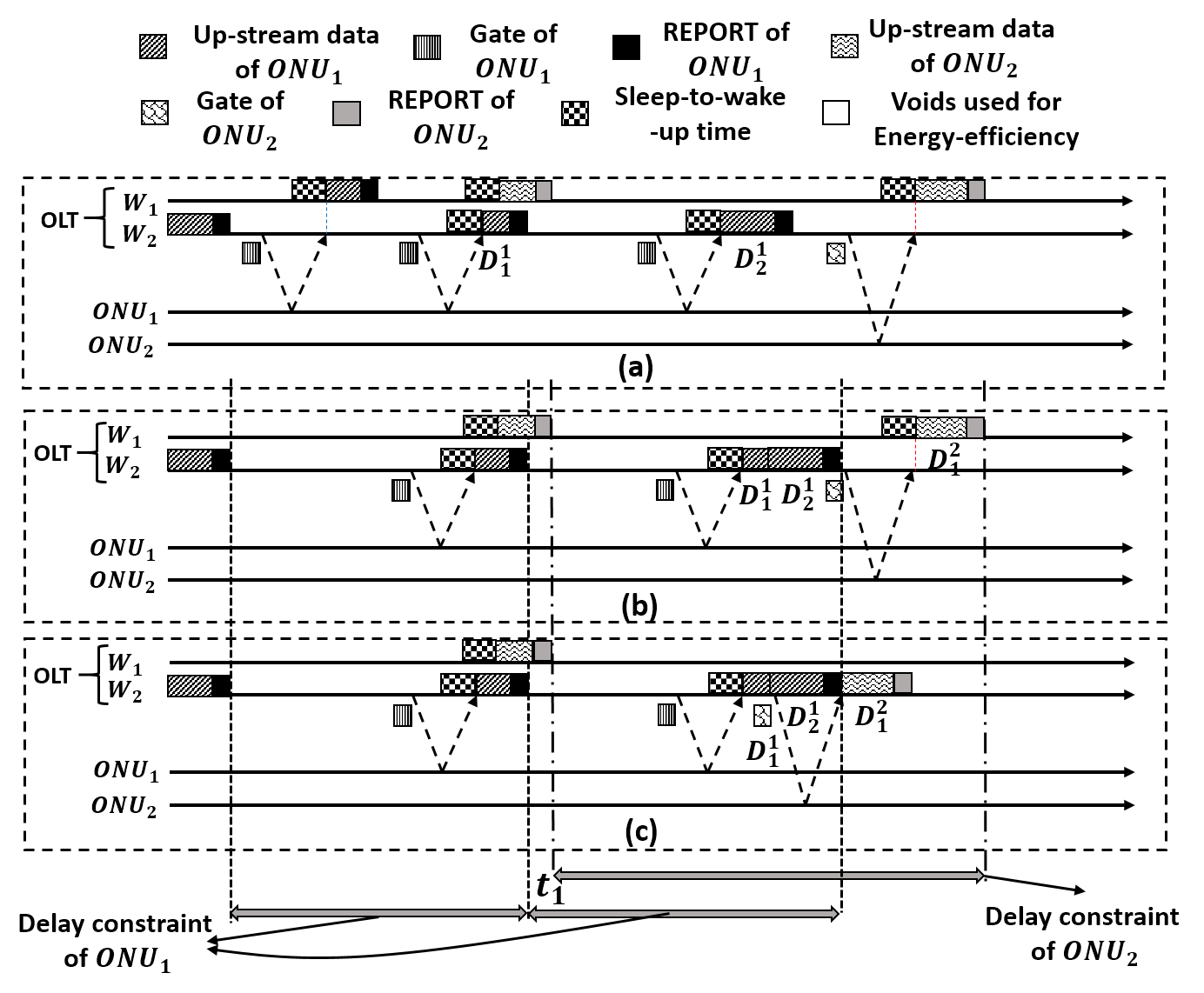}
		\caption{Motivation of the proposed EO-NoVM protocol}
		\label{ill}
	\end{figure}

	In Fig. \ref{ill}, we illustrate the methodology used for the proposed EO-NoVM algorithm. Three different 
	scheduling mechanism is shown in Fig. \ref{ill}. In Fig. \ref{ill}(a), an arbitrary scheduling process is considered. The scheduling process considered in Fig. \ref{ill}(b) is focused towards minimization of the number of REPORT messages, realized by scheduling to the latest time instant while maintaining the delay constraint. It can be observed from Fig. \ref{ill}(a) and Fig. \ref{ill}(b) that the total duration over which the up-stream data transmitted by $ONU_1$ and $ONU_2$ is unchanged by the scheduling process. Further, the first GATE message of $ONU_1$ is delayed in Fig. \ref{ill}(b) as compared to Fig. \ref{ill}(a). As a consequence, $D_1^1$ and $D_2^1$ is clubbed in the second case. This not only reduces the number of REPORT messages by one but also reduces the loss due to sleep-to-wake-up time. Hence, the effective duration over which energy-efficiency can be achieved is improved in the second case (Fig. \ref{ill}(b)) as compared to first case (Fig. \ref{ill}(a)).
	
	The scheduling process considered in Fig. \ref{ill}(c) is same as the methodology used for our proposed EO-NoVM algorithm. Let us assume that scheduling up-to $t_1$ is already performed. At $t_1$, the REPORT message from $ONU_2$ is arrived. If the scheduling is focused towards minimization of the number of REPORT messages then $D_1^2$ is scheduled at the delay boundary (Fig. \ref{ill}(c)). In this case, scheduling of $D_1^2$ can be performed in any of the two wavelengths. In both of the cases, one extra void is created by scheduling process. While, in our proposed mechanism, this extra void creation can be avoided by scheduling $D_1^2$ with the transmission slot of $ONU_1$ (Fig. \ref{ill}(c)). However, $D_1^2$ can be scheduled either before or after the transmission slot of $ONU_1$. As we have already discussed, if multiple clubbing opportunity is present then DWBA is focused towards scheduling as late as possible. Hence, $D_1^2$ is scheduled after the transmission slot of $ONU_1$ (Fig. \ref{ill}(c)). Further, it can be observed from Fig. \ref{ill} that the effective duration over which energy-efficiency can be achieved is maximum in the third case (Fig. \ref{ill}(c)).   
	\subsection{EO-NoVM}
\begin{figure}[h]
		\centering
		\includegraphics[scale=0.4]{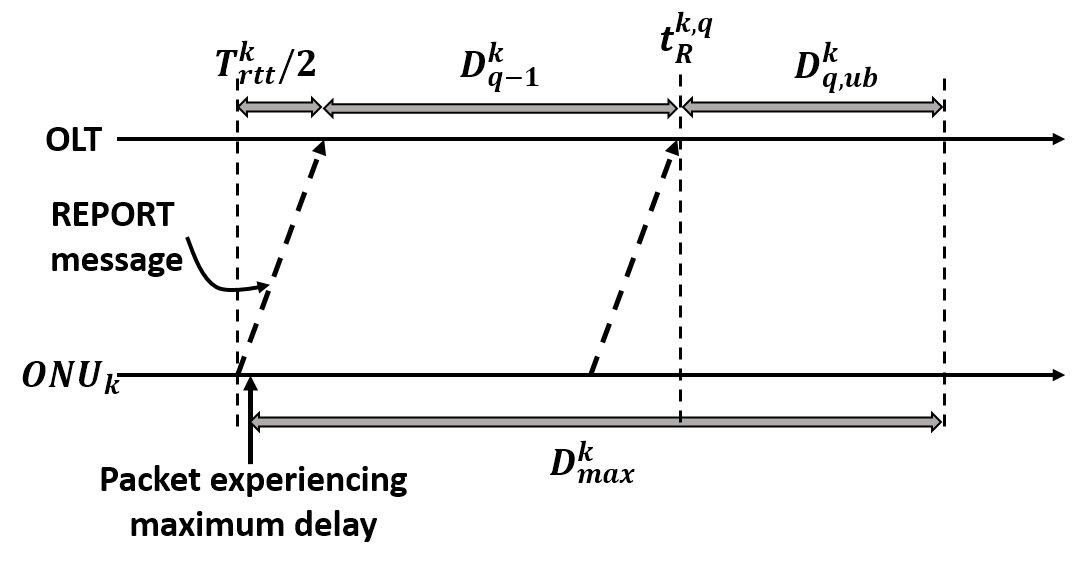}
		\caption{Delay Calculation}
		\label{del}
	\end{figure}

	In this subsection, we discuss the proposed EO-NoVM algorithm. For describing the EO-NoVM protocol, we assume that the OLT connects to $N$ ONUs and $W$ number of wavelengths are present at the OLT for scheduling. Further, the time required by a tunable laser (at ONU) to tune from wavelength $i$ to $j$ is assumed to be $T_t^{i,j}$. In this paper, we have assumed; $T_t^{i,j}=|i-j|$ $T_t$ where $T_t$ is the tuning time between two consecutive wavelengths.  Moreover, the currently tuned wavelength of $ONU_k$ is assumed to be $w_c^k$. Let at time $t_R^{k,q}$, the $q^{th}$ REPORT message from the $k^{th}$ ONU ($ONU_k$) reaches the OLT. The protocol followed by the OLT at $t_R^{k,q}$ to schedule $ONU_k$ is discussed below. 
	Since, the REPORT message from $ONU_k$ \st {is} arrived \st {to} the OLT at $t_R^{k,q}$, the minimum time instant at which the GATE massage for $ONU_k$ can be transmitted by the OLT is $t_R^{k,q}+T_p$ where $T_p$ is the GATE message generation time. The time required by $ONU_k$ to tune its transmitter to wavelength $j$ after receiving the GATE message is given by $T_t^{w_c^k,j}$. Thus there exists a minimum time instant ($TC_{min}^{k,j}$) at which the first bit of upstream data of $ONU_k$ reaches the OLT if it is scheduled at wavelength $j$ (Fig. \ref{del}). If $T_{tx}$ and $T_{rtt}^k$ are the GATE message transmission time and round-trip time of $ONU_k$ respectively then 〖$TC_{min}^{k,j}$ given by \ref{min1}.
	\begin{align} \label{min1}
	TC_{min}^{k,j} (\forall j)=t_R^{k,q}+T_p+T_{tx}+T_{rtt}^k+T_t^{w_c^k,j}
	\end{align}
	\begin{figure}[b]
		\centering
		\includegraphics[scale=0.3]{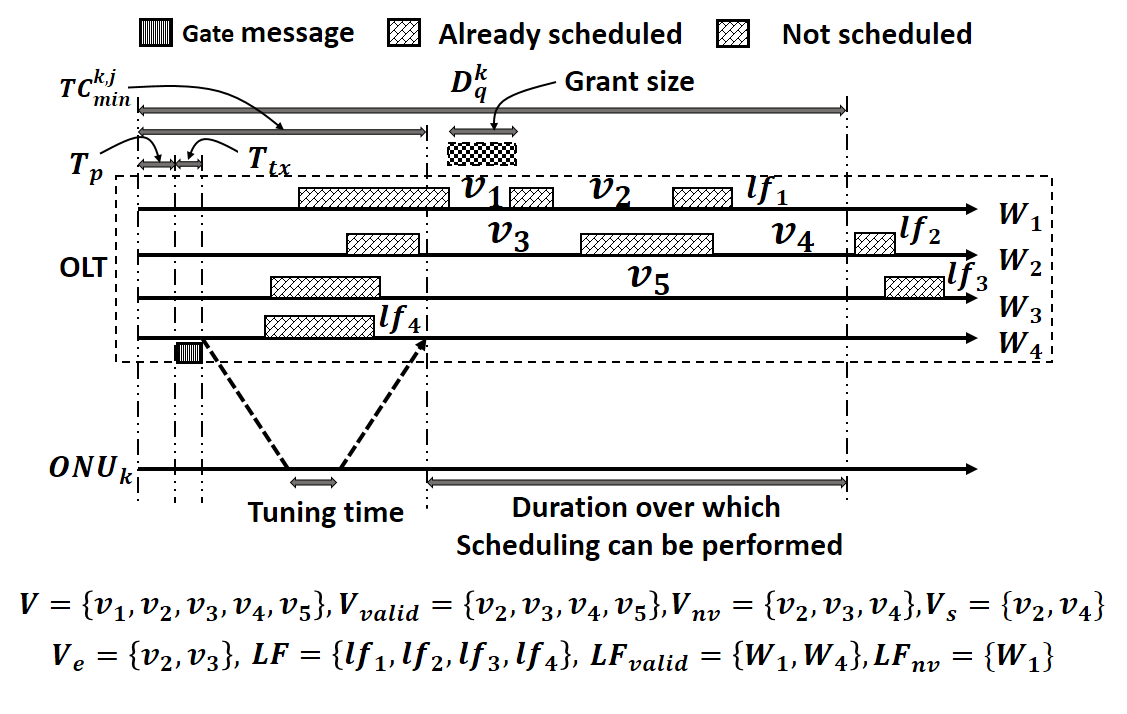}
		\caption{Illustration of the proposed EO-NoVM protocol (assuming tuning time for all wavelengths same)}
		\label{ill1}
	\end{figure} 
	
	In this paper, we assume that all packets of $ONU_k$ belong to the same class of traffic, and therefore, they have constant QoS delay constraint $D_{max}^k$ (delay is defined as time interval between the packet arrival at the OLT to packet arrival to the ONU). Further, in this paper, the grant-sizing scheme \cite{surve1,mcgarry2004ethernet} is assumed to follow the gated scheme \cite{surve1}. However, the protocol can be extended even for the other grant-sizing schemes as well.  In the gated scheme, the granted bandwidth of an ONU is exactly equal to the bandwidth requested by that ONU through the previous REPORT message. Therefore, in the gated scheme, the packet that has arrived to the ONU immediately after sending the previous REPORT message faces the maximum delay. If the time difference between the $(q-1)^{th}$ and the $q^{th}$ REPORT message of $ONU_k$ is $D_q^k$ then the upper limit of the time interval between the $q^{th}$ and the $(q+1)^{th}$ REPORT message for maintaining the delay constraint is given by \ref{dela}.
	\begin{align}\label{dela}
	D_{max}^k=D_{q-1}^k+D_q^k+\frac{T_{rtt}^k}{2}
	\end{align}
	 
	If $D_q^k\leq D_{const}^k (\forall q)$ then from \ref{dela}, $D_{max}^k\leq 2D_{const}^k+T_{rtt}^k/2$. If we wish to satisfy the delay constraint while keeping $D_q^k$ fixed $\forall q$ (independent of $D_{q-1}^k$) then $D_q^k$ can be calculated from $D_{max}^k$ by \ref{cons}.
	\begin{align}\label{cons}
	D_q^k (\forall q)=D_{const}^k=\frac{D_{max}^k-T_{rtt}^k/2}{2}
	\end{align}
If the calculation $D_q^k$ from $D_{max}^k$ follows \ref{dela} then we define it as variable delay-bound calculation whereas if the same follows \ref{cons} then it is defined as fixed delay-bound calculation. However, it can be noted that satisfaction of delay constraint by employing fixed delay-bound calculation requires $D_{q-1}^k\leq D_{const}^k$. In this paper, fixed delay-bound calculation is  used for calculating $D_q^k$ from $D_{max}^k$ (refer \ref{ssec:compa} for further discussion) as long as $D_{q-1}^k\leq D_{const}^k$.  
 
 Following the previous arguments, the scheduling of $ONU_k$ at wavelength $j$ can only be performed in between $TC_{min}^{k,q}$ and $t_R^{k,q}+D_q^k$ while satisfying the delay constraint. We observe that two possible ways are present to perform the scheduling of $ONU_k$ within this interval.
\begin{itemize}
	\item Schedule within the existing voids (void filling)
	\item Schedule after the latest scheduling horizon (the latest time instant up to which scheduling has already been performed of wavelengths.
\end{itemize}

In EO-NoVM, void filling is preferred over scheduling after the latest scheduling horizon. This is because, void filling prevents extension of the scheduling horizon and therefore, might allow to schedule some other ONUs after the latest scheduling horizon while satisfying their delay constraint. Below we discuss the exploitation of this two methods in EO-NoVM algorithm. 

From the above discussion, the basic methodology used by the OLT to schedule $ONU_k$ in EO-NoVM protocol can be summarized as follows:
\begin{itemize}
	\item OLT first try to combine the US data of $ONU_k$ with already scheduled US data
	\item If the clubbing of US data is possible both by void filling as well as scheduling after the latest scheduling horizon then void filling is preferred
	\item If the previous step provides multiple clubbing opportunities then OLT try to schedule $ONU_k$ as late as possible
	\item If consolidation of US data is not possible then again the previous two steps are followed 
\end{itemize}
Below we describe the proposed EO-NoVM protocol in detail.
\begin{figure}[t]
	\centering
	\includegraphics[scale=0.4]{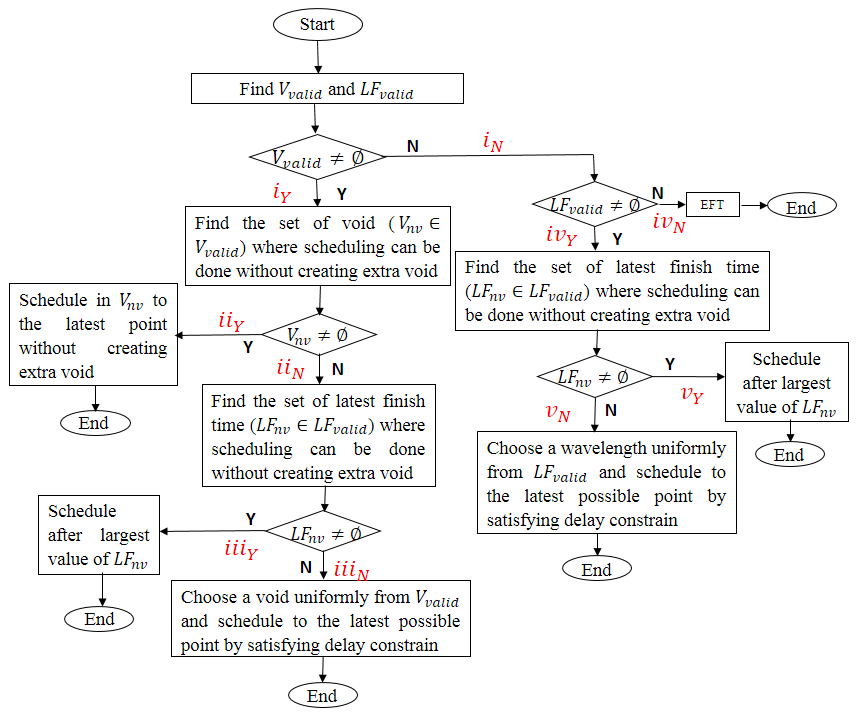}
	\caption{Flowchart of EO-NoVM}
	\label{flw}
\end{figure} 
~~Let us assume that $V$ is the set of all voids. The number of elements of $V$ is upper-limited by $N$ \cite{eft}. Therefore, $V$ can be written as; $V=\{v_1,v_2,...,v_p\}$ where $v_i$ denotes the $i^{th}$ void. Each void $v_i$  has three components; start time ($v_i^s$), end time ($v_i^e$) and wavelength ($v_i^w$). Further, we assume $LF$ is a set of the latest scheduling horizon of all wavelengths, and hence, $LF$ consists of $W$ elements. $LF$ can be written as; $LF=\{lf_1,lf_2,...,lf_W\}$ where $lf_j$ denotes the latest scheduling horizon of the $j^{th}$ wavelength. 

In EO-NoVM, OLT first find the set of voids within which $ONU_k$ can be scheduled while satisfying the delay constraint. This is possible if the size of the void ($v_i^e-v_i^s$) is bigger than US transmission window ($T_w^k$) of $ONU_k$ consists of US data transmission of size $G_k$ (Grant-size), REPORT message transmission and guard duration. Therefore, $T_w^k$ is given by; $$T_w^k=G^k T+N_R T+T_g$$However, the possibility of $TC_{min}^{k,j}>v_i^s$ ($v_3 \text{ and }v_5$ in Fig. \ref{ill1}) and $v_i^e>t_R^{k,q}+D^k_q$ ($v_4 \text{ and }v_5$ in Fig. \ref{ill1}) further restrict the effective width of the void $v_i$ ($\forall i$ that can be utilized for scheduling $ONU_k$. Hence, $V_{valid}$ (Fig. \ref{ill1}) can be calculated by \ref{Vv}.
\begin{align}\label{Vv}
\nonumber
V_{valid}=\{i|\min⁡(v_i^e,t_R^{k,q}+D_{q}^k)-\max⁡(〖v_i^s,TC〗_{min}^{k,j} )\\ \geq T_w^k  \text{~~}  \forall i\in arg{(V)} \}
\end{align}

This is followed by the calculation of the set of wavelengths where scheduling of $ONU_k$ can be performed after the latest scheduling horizon by maintaining the specified delay bound ($LF_{valid}$). By using the same logic used for calculating \ref{Vv}, $LF_{valid}$ (Fig. \ref{ill1}) is given by \ref{LFv}.
\begin{align}\label{LFv}
\nonumber
LF_{valid}=\{j|t_R^{k,q}+D_{q}^k-\max⁡(〖lf_j,TC〗_{min}^{k,j} )\\ \geq T_w^k   \text{~~}  \forall j\in arg{(LF)} \}
\end{align}
\subsection*{Case 1: $V_{valid}=LF_{valid}=\Phi$}
If $V_{valid}=LF_{valid}=\Phi$ ($iv_N$ in Fig. \ref{flw}) then the delay constraint of $ONU_k$'s US data might get violated. Therefore, in this case, $ONU_k$ is required to be scheduled as early as possible, and hence, EFT \cite{eft} is used for scheduling in EO-NoVM.  Since, $ONU_k$ is scheduled after the delay limit ($D_{const}^k$),   $D_{q}^k> D_{const}^k$. Therefore, the calculation of $D_{q+1}^k$ requires employing variable delay-bound calculation (\ref{dela}). For the rest of the cases, $D_{q+1}^k$ is calculated by using fixed delay-bound calculation (\ref{cons}). 
\subsection*{Case 2: $V_{valid}\neq \Phi \text{ or } LF_{valid}\neq \Phi$}
If $V_{valid}\neq \Phi$ or $LF_{valid}\neq\Phi$ then the delay constraint of $ONU_k$'s US data  
is guaranteed to be satisfied. After guaranteeing the delay constant, OLT focuses on scheduling $ONU_k$ by maximizing its energy-efficiency. In this process, clubbing of US data is employed in EO-NoVM algorithm (\ref{ssec:moti}). Since void filling is preferred over scheduling after the latest scheduling horizon in our proposed protocol, OLT first check $V_{valid}=\Phi$ or not.
\subsubsection*{$\mathbf{V_{valid}\neq \Phi}$}
If $V_{valid}\neq \Phi$ ($i_Y$ in Fig. \ref{flw}) then the possibility of US data clubbing within those voids ($\in V_{valid}$) are verified. In this process, OLT finds the set of voids ($V_{nv}\in V_{valid}$) where $ONU_k$'s US data can be combined with the other ONUs US data that have already scheduled (Fig. \ref{ill1}). The consolidation of the US data within the voids can be achieved in two possible ways:  
\begin{itemize}
	\item Initialization of $ONU_k$'s US data transmission exactly at the start time of the void
	\item Termination of $ONU_k$'s US data transmission window ($T_w^k$) immediately before the end time of the void
\end{itemize} 
Depending on this two possibilities, EO-NoVM scheduler finds another two sets $V_s$ and $V_e$ where $V_s$ and $V_e$ ($\in V_{valid}$) represents the set of voids where US data clubbing is possible at the start of the voids and the end of the void respectively. Therefore, a void ($v_i$) is included into set $V_s$ if $v_i^s\geq TC_{min}^{k,v_i^w}$ while it is included in set $V_e$ if $v_i^e\leq t_R^{k.q}+D_k^q$. Hence, $V_s$ and $V_e$ can be calculated by \ref{VS} and \ref{VE} respectively.
\begin{align} \label{VS}
V_s=\{i|v_i^s\geq TC_{min}^{k,v_i^w} \text{~~} i\in V_{valid}\}
\end{align}  
\begin{align}\label{VE}
V_e=\{i|v_i^e\leq t_R^{k.q}+D_k^q \text{~~} i \in V_{valid}\}
\end{align}   

If $V_{nv}\neq \Phi$ ($ii_Y$ in Fig. \ref{flw}) which is possible if $V_s \neq \Phi$ or $V_e \neq \Phi$ then consolidation of US data of $ONU_k$ within the void is possible while maintaining the delay constraint. Therefore, in this case, scheduling of $ONU_k$ is performed within the void ($\in V_{nv}$) to the latest possible time instant (\ref{ssec:moti}). To facilitate this, EO-NoVM scheduler finds the void with the maximum start-time ($v_{sm}$) from $V_s$ and the void with maximum end-time ($v_{em}$) from $V_e$. If $v_{v_{sm}}^s+T_w^k≥v_{v_{em}}^e$ then 〖$ONU〗_k $ is scheduled in such a way that the first byte of upstream data reaches to the OLT from 〖$ONU〗_k$ exactly at $v_{sm}^s$. In this case, the time instance of generation of GATE message for 〖$ONU〗_k $ ($T_{sch}^k$) and the scheduled wavelength of 〖$ONU〗_k $ ($W_{sch}^k$) is given by \ref{W} and \ref{T} respectively.    
\begin{align}\label{W}
W_{sch}^k=v_{v_{sm}}^w
\end{align}
\begin{align}\label{T}
T_{sch}^k=v_{v_{sm}}^s-TC_{min}^{k,W_{sch}^k}
\end{align}

Otherwise,  〖$ONU〗_k $is scheduled in such a way that the guard duration ends immediately before the $v_i^e$. In this case, $T_{sch}^k$ and $W_{sch}^k$ is given by \ref{We} and \ref{Te} respectively.
\begin{align}\label{We}
W_{sch}^k=v_{v_{em}}^w
\end{align}
\begin{align}\label{Te}
T_{sch}^k=v_{v_{em}}^s-TC_{min}^{k,W_{sch}^k}-T_w^k
\end{align}

If $V_{nv}=\Phi$ ($ii_N$ in Fig. \ref{flw}) then void filling while combining the $ONU_k$'s US data with the already scheduled US data is not possible. However, the clubbing of US data can further be achieved by scheduling $ONU_k$ after the latest scheduling horizon of some wavelengths. In this process, OLT finds a set of wavelengths ($LF_{nv} \in LF_{valid}$) at which scheduling of $ONU_k$ can be performed after the latest scheduling horizon while achieving the US data clubbing (Fig. \ref{ill1}). This is possible only if $lf_j\geq TC_{min}^{k,j}$ . Hence, $LF_{nv}$ can be given by \ref{lnv}.
\begin{align}\label{lnv}
LF_{nv}=\{j|lf_j\geq TC_{min}^{k,j} \text{~~} \forall j \in LF_{valid} \}
\end{align}

If $LF_{nv}\neq \Phi$ ($iii_Y$ in Fig. \ref{flw}) then merging of $ONU_k$'s US data can be achieved by scheduling it after the latest scheduling horizon. After guaranteeing US data clubbing, the scheduling $ONU_k$ is directed towards the latest possible time instance (\ref{ssec:moti}). For doing so, OLT finds the wavelength with height value of latest scheduling horizon ($F_m$) from $LF_{nv}$. Hence, in this case, $T_{sch}^k$ and $W_{sch}^k$ is given by \ref{Wfc} and \ref{Tfc} respectively.
\begin{align}\label{Wfc}
W_{sch}^k=F_m \text{~~where~~} F_m=\max_{j\in LF_{nv}}lf_j
\end{align}
\begin{align}\label{Tfc}
T_{sch}^k=lf_{F_m}-TC_{min}^{k, F_m}
\end{align}

If $LF_{nv}=\Phi$ ($iii_N$ in Fig. \ref{flw}) then the possibility of clubbing of $ONU_k$'s US data does not exists. Therefore, one extra void is definitely created by scheduling. Since, in this case, $V_{valid\neq \Phi}$, void filling is possible by satisfying the delay constraint is possible. It can be noted that the possibility of scheduling to the latest possible point exists in all of the voids ($\in V_{valid}$) as $V_{nv}=\Phi$ ensures $v_i^e> t_R^{k.q}+D_k^q$ $ \forall$$ v_i$ $\in V_{valid}$. Therefore, scheduling of $ONU_k$ is performed to any of the voids ($\in V_{valid}$) to the latest possible time instant while maintaining the delay constraint. Hence, in this case, $T_{sch}^k$ and $W_{sch}^k$ is calculated by \ref{Wfv} and \ref{Tfv} respectively.  
\begin{align}\label{Wfv}
W_{sch}^k=v_{temp}^w \text{~~where~~} temp=uniform(V_{valid})
\end{align}
\begin{align}\label{Tfv}
T_{sch}^k=t_R^{k.q}+D_k^q-T_w^k-TC_{min}^{k,W_{sch}^k}
\end{align}

\subsubsection{$\mathbf{V_{valid}=\Phi}$}
In this case ($iv_Y$ in Fig. \ref{flw}), void-filling while satisfying the delay constraint is not possible. However, $LF_{valid}\neq \Phi$ ensures scheduling $ONU_k$ after the latest scheduling horizon of wavelengths ($\in LF_{valid}$) by by fulfilling the delay constraint. In this case also, clubbing of US data of $ONU_k$ is preferred. For doing so $LF_{nv}$ is calculated by using \ref{lnv}. If $LF_{nv}\neq \Phi$ ($v_Y$ in Fig. \ref{flw}) then $W_{sch}^k$ and $T_{sch}^k$ are calculated by \ref{Wfc} and \ref{Tfc} respectively. Otherwise ($v_N$ in Fig. \ref{flw}), scheduling of $ONU_k$ is performed in any of the wavelength ($\in LF_{valid}$) to the latest possible time instance. Hence, in this case, $T_{sch}^k$ and $W_{sch}^k$ is given by \ref{Wfv1} and \ref{Tfv1} respectively. 
\begin{align}\label{Wfv1}
W_{sch}^k=uniform(LF_{valid})
\end{align}
\begin{align}\label{Tfv1}
T_{sch}^k=t_R^{k.q}+D_k^q-T_w^k-TC_{min}^{k,W_{sch}^k}
\end{align}
\begin{figure}[!t]
	\removelatexerror
 \begin{algorithm}[H]
 	\SetAlgoLined
 	\caption{EO-NoVM}
 	\KwIn{$V$, $LF$, $D_{max}^k$}
 	\KwOut{$W_{sch}^k$, $T_{sch}^k$}
 	\Parameter{$V_{valid}$, $LF_{valid}$, $V_{nv}$, $LF_{nv}$, $V_s$, $V_e$	
 	}
 	\BlankLine
 	\Begin{
 		Calculate $V_{valid}$ and $LF_{valid}$ from $V$ and $LF$ by using \ref{Vv} and \ref{LFv}.\\
 		\eIf{$V_{valid}\neq\Phi$}{
 		Calculate $V_s$ and $V_e$ using \ref{VS} and \ref{VE} respectively\\
 		\uIf{$V_s\neq \Phi$ or $V_e \neq \Phi$}
 		{\If{$V_s\neq \Phi$}{$v_{sm}=arg(\max_{i\in V_s}v_i^s)$}
 		\If{$V_e\neq \Phi$}{$v_{em}=arg(\max_{i\in V_e}v_i^s)$}	
 		\uIf{$V_s\neq \Phi$ and $V_e\neq \Phi$}{
 			\eIf{$v_{v_{sm}}^s+T_w^k>v_{v_{em}}^e$}{Calculate $W_{sch}^k$ and $T_{sch}^k$ by using \ref{W} and \ref {T}}{Calculate $W_{sch}^k$ and $T_{sch}^k$ by using \ref{We} and \ref {Te}}
 		}
 	 	\uElseIf{$V_s\neq \Phi$}{Calculate $W_{sch}^k$ and $T_{sch}^k$ by using \ref{W} and \ref {T}}
 	 	\Else{Calculate $W_{sch}^k$ and $T_{sch}^k$ by using \ref{We} and \ref {Te}}
 	}
 \Else{ Calculate $LF_{nv}$ by using \ref{lnv}	\\
 	\eIf{$LF_{nv}\neq \Phi$}{Calculate $W_{sch}^k$ and $T_{sch}^k$ by using \ref{Wfc} and \ref{Tfc}}{Calculate $W_{sch}^k$ and $T_{sch}^k$ by using \ref{Wfv1} and \ref{Tfv1}}}
}{\eIf{$LF_{valid}\neq \Phi$}{Calculate $LF_{nv}$ by using \ref{lnv}\\ \eIf{$LF_{nv}\neq \Phi$}{Calculate $W_{sch}^k$ and $T_{sch}^k$ by using \ref{Wfc} and \ref{Tfc}}{Calculate $W_{sch}^k$ and $T_{sch}^k$ by using \ref{Wfv1} and \ref{Tfv1}}
}{Follow EFT \cite{eft} for scheduling}}
}\end{algorithm}
\end{figure}
\section{Complexity analysis}\label{comp}
In this section, we analyze the algorithmic complexity of EO-NoVM. For protocols incorporating online scheduling, online complexity (time taken to schedule ONUs after receiving a REPORT message) is generally considered. Since calculation of $V_{valid}$ requires a search through all the $N$ voids (\ref{Vv}), the complexity of calculating $V_{valid}$ is $O(N)$. $V_s$ (\ref{VS}) and $V_e$ (\ref{VE}) can similarly be calculated with the same complexity. Since $V$ and $LF$ are updated only after the scheduling process ends, both $V$ and $LF$ can be kept sorted before receiving the REPORT message. Thus, the complexity of sorting does not affect the complexity of the online scheduling process. For different $ONU_k$, $LF$ ($lf_j \forall j$) can be kept sorted (at the OLT scheduler, in an offline process) with respect to the value of $lf_j-T_t^{w_c^k,j} \forall k$, as both $w_c^k$ and $lf_j$ are known to the OLT even before receiving the REPORT message from $ONU_k$. Therefore, binary search can be used for calculating $\max(lf_j, TC_{min}^{k,j})$ in (\ref{LFv}) which has a complexity of $O(log⁡(W))$. Now, \ref{LFv} can be re-written by \ref{LFv1}. 
\begin{align}\label{LFv1}
LF_{valid}=\{j|\max(lf_j, TC_{min}^{k,j}) \geq t_R^{k,q}+D_{q}^k-T_w^k\}  
\end{align}
The right hand side of \ref{LFv1} become constant after receiving the REPORT message. Therefore, binary search can be employed for calculating $LF_{valid}$, and it provides a complexity of $O(log⁡(W))$. Similarly, $LF_{nv}$ (\ref{lnv}) can be calculated with a complexity of $O(log⁡(W))$. If $LF$ is kept sorted with respect to the value of $lf_j$, then $F_m$ (\ref{Wfc}) can further be calculated with a complexity of $O(log⁡(W))$. Therefore, if $N+1$ copies of $LF$ is maintained at the OLT where $N$ copies are kept shorted with respect to the values of $lf_j-T_t^{w_c^k,j}$ for different $ONU_k$, while the other one is kept sorted in terms of $lf_j$, then the overall online complexity of EO-NoVM is given by; $O(N+log⁡(W))$.
\section{Upper bound of energy-efficiency calculation}\label{maxcal}
In this section, we calculate the maximum energy-efficiency figures attainable. The best energy-efficiency figure is achieved if all active periods of the OLT receiver can be clubbed together. Under such scenarios, no voids are created resulting in no loss due to $T_sw$. However, it is very difficult to achieve this maximum limit of energy-efficiency due to the presence of possibility of void creation due to scheduling.  
We assume  $r_a^m$ be the maximum arrival rate and L (0≤L≤1) be the fractional load of any of the ONUs. Then the average arrival rate of the individual ONUs is given by $r_a^m L$ while the total arrival rate of all $N$ ONUs is $〖Nr〗_a^m L. $.  The overall rate of departure from all ONUs is $Wr_d$ (\ref{ssec:moti}). The active period of the OLT receiver ($T_{OLT}$) is given by; $$T_{OLT}=\frac{Nr_a^mL}{Wr_d}$$
Consequently, the fraction of inactive period is; ($1-T_{OLT}$). Since, there is no loss due to $T_sw$ in the scenario of maximum energy-efficiency, the maximum energy-efficiency ($\eta_{max}$) can be given by \ref{etamax}.
\begin{align}\label{etamax}
\eta_{max}=\Big(1-\frac{Nr_a^mL}{Wr_d}\Big)100\%
\end{align}
\section{Results and discussions}\label{rd}
In this section, we evaluate the performance of the proposed EO-NoVM algorithm in terms of energy-efficiency (evaluated with respect to the traditional OLT receivers power consumption) and average delay. In the first subsection, the effect of variable and fixed delay constraint calculation on energy-efficiency is demonstrated. The energy-efficiency figures achieved by employing EO-NoVM algorithm in EPON is compared with the same achieved by employing the conventional approach of minimization of the number of active wavelengths in the next subsection. This is followed by a discussion on the effect of different parameters on energy-efficiency and average delay in the next subsection. 
The performance evaluation of EO-NoVM is performed with the help simulation results produced in OMNET++. The network run time for the simulation has been considered to be 20s. The traffic arrival process at every ONU is assumed to be self-similar \cite{self}. We have also assumed that the ON and OFF periods of the self-similar traffic follows Pareto distribution with shape parameters $\alpha_{ON}=1.2$ and $\alpha_{OFF}=1.4$ ($\alpha_{ON}$-shape parameter of the ON period, $\alpha_{OFF}$-shape parameter for the OFF period).   
\begin{table}[h]
	\caption{Parameter used for simulation}
	\begin{center}
		\begin{tabular}{|c|c|}
			\hline
			\multicolumn{1}{|c|}{\textbf{Parameters}} & 
			\multicolumn{1}{c|}{\textbf{Value}}\\ \hline
			
			$r_a^m$ & $100 Mbps$ \\
			$r_d$ & $1 Gbps$\\
			$T_p$ & $35ns$\\
			$T_g$ & $5\mu s$\\
			$T_{tx}$ & $512 ns$\\ 
			$T_t$ & $1 \mu s$\\
			$N_R$ & $64 Bytes$\\
			\hline
		\end{tabular}
	\end{center}
\end{table}
\subsection{Effect of variable and fixed delay constraint calculation on energy-efficiency}\label{ssec:compa}
\begin{figure}[h]
	\centering
	\includegraphics[scale=.7]{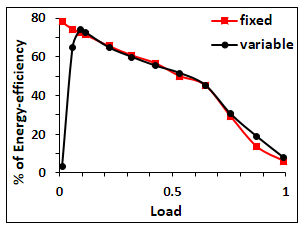}
	\caption{Effect of variable and fixed delay constraint calculation on energy-efficiency}
	\label{delay}
\end{figure}

There exists two different ways of calculation of $D_q^k$ from $D_{max}^k$ using \ref{dela} and \ref{cons} as shown in Section II(b). In this subsection, we study the effect of these two procedure on energy-efficiency. For this purpose, energy-efficiency as a function of load for two different ways of calculation of $D_q^k$ is plotted in Fig. \ref{delay} (we have considered a scenario with $N=16$, $ W=2$, $T_{rtt}^k=0.2ms (\forall k)$, $T_{sw}=2ms$, $D_{max}^k (\forall k)=10ms$. If $D_q^k$ is calculated from \ref{dela} then $D_q^k$ varies with $q$ and is dependent on $D_{q-1}^k$. It is observed from Fig. \ref{delay} that below a certain load ($\sim 0.08$), the energy-efficiency for variable $D_q^k$ increases with increase of load. The reason behind this observation is explained as follows. If $D_q^k\leq T_{min}^{k,j}~\forall k$ (i.e., delay constraint can never be satisfied) then, according to our protocol the scheduler follows EFT. Therefore, all the wavelengths are used for scheduling and hence creates one extra void by each scheduling in most of the cases. In that case, $D_{q+1}^k$ attains a very high value compared to $D_q^k$ (refer \ref{dela}).  However, the grant-size of all the ONUs are very small at low load scenario. As such, the latest scheduling horizon of all wavelengths is smaller than $T_{min}^{k,j} \forall k$ in most of the cases. Hence, extra voids are created by scheduling ONUs to the furthest time possible (limited by the delay constraint).  Therefore,$ D_{q+2}^k$ will then attain a comparatively lower value with respect to $D_{q+1}^k$ and there is a high possibility that the scheduler again follows EFT [20]. Following a similar argument, we can show that $D_q^k$ will alternate between a high and a low value for subsequent values of q to create a large number of voids.  With an increase in the network load, the grant-size also increases and the probability that the latest scheduling horizon happens later than 〖$T_{min}^{k,j}$ increases. As a consequence, the number of voids created decreases with an increase in the network load. After a certain load ($\sim 0.08$), the scheduler gets opportunity to club upstream data transmission of most of the ONUs. Therefore, the variation of $D_q^k$ is significantly reduced making the energy-efficiency similar to that of a fixed $D_q^k$. Whereas, we observe that employing a fixed $D_q^k$ allows significant improvement in the energy-efficiency than adopting variable $D_q^k$ at low traffic load conditions. Therefore, in EO-NoVM we have chosen a fixed $D_q^k$.
It is observed that the variable $D_q^k$ provides a small increment of energy-efficiency at high load (for $N=16,W=2$). Therefore, it is possible to design an algorithm with adaptive $D_q^k$ calculation (i.e. fixed $D_q^k$ at low load and variable $D_q^k$ at high load). This might provide further improvement in energy-efficiency compared to EO-NoVM. However, this requires efficient traffic prediction mechanism which eventually makes the protocol more complicated. 
\subsection{Comparison of EO-NoVM and the number of active minimization schemes}\label{ssec:compb}
In this subsection, the proposed EO-NoVM algorithm is compared with the convention approach of minimization of the number of active wavelengths in terms of energy-efficiency. All the existing proposal for energy-efficient OLT design in TWDM-PON is suitable for offline scheduling except the protocol proposed in \cite{online}. However, it is well known that online scheduling is essential for long reach PON with stringent delay constraint especially at high load scenarios. Therefore, in this paper, the comparison is made with the only proposed online scheduling scheme employing the minimization of the number of active wavelengths \cite{online} (WM-\cite{online}).      
\begin{figure}[h]
	\centering
	\includegraphics[scale=.7]{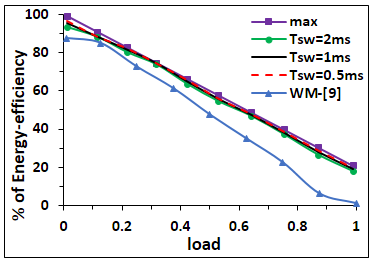}
	\caption{Comparison of EO-NoVM with the conventional approach of minimization of the number of active wavelengths for different values of $T_{sw}$}
	\label{compi}
\end{figure} 
 \begin{figure*}[t]
	\centering
	\includegraphics[scale=.75]{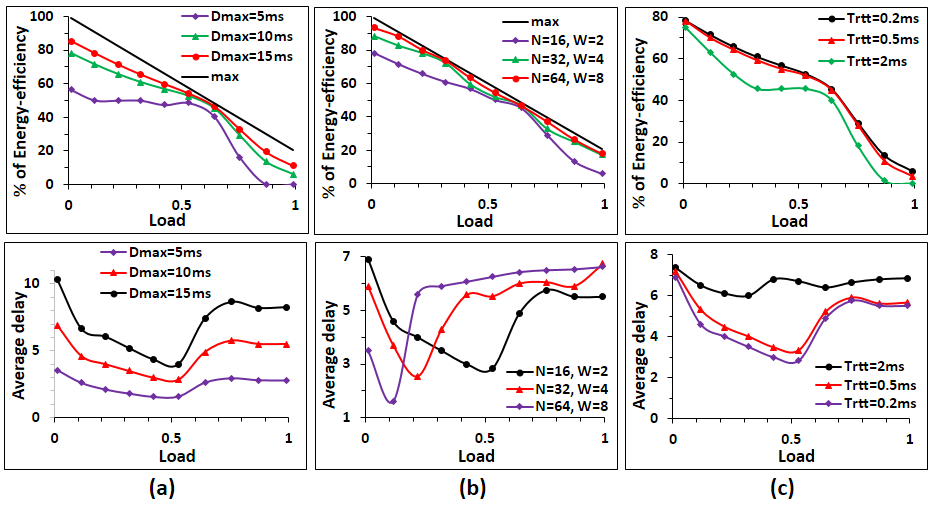}
	\caption{Effect of (a) delay constraint (b) scaling and (c) round-trip time on the achieved energy-efficiency and the average delay in the proposed EO-NoVM algorithm. D{max}-$D_{max}^k\forall k$, max-theoretical upper bound}
	\label{effect}
\end{figure*}
 Energy-efficiency as a function of the traffic load for different values of $T_{sw} (=0.5ms,1ms,2ms)$ is plotted in Fig. \ref{compi}. We choose the same parameters as mentioned in the previous sub-section (Fig. \ref{delay}) except for $N=64$, $W=8$. In Fig. \ref{compi}, the theoretically calculated upper bound of energy-efficiency (\ref{etamax}) is also plotted against the network load to illustrate the best attainable performance of energy-efficiency figures. Further, energy-efficiency against the offered load for WM-\cite{online} is plotted in Fig. \ref{compi} (for comparison). It can be observed from Fig. \ref{compi} that at very low load, the deviation of the achieved energy-efficiency by using WM-\cite{online} from the theoretical upper bound decreases with an increase in the traffic load. The reason behind this can be explained as follows. The schemes that employs the minimization of the number of active wavelengths requires to keep at least one wavelength always active. As a consequence, at very low load, a lot of voids are created within the active wavelength due of unavailability of up-stream data which remain unutilize for energy-efficiency. The increment of traffic load leads to creation of lesser amount of void within the active wavelength due to increment effective utilization factor ($\rho$) and hence, the achieved energy-efficiency from WM-\cite{online} reaches closer to the theoretical upper bound with an increase in the traffic load. Moreover, it can be observed that beyond a certain load, this deviation starts increasing with the increment of traffic load. This is because,  the number of active wavelengths and hence, the amount of voids created within the active wavelengths increases with an increase in the traffic load.
 
 However, this loss of energy-efficiency due to the presence of the untilized voids within the active wavelengths is absent in our proposed EO-NoVM algorithm. This results in a significant improvement of energy-efficiency as compared to WM-\cite{online}. Further, the achieved energy-efficiency by employing the proposed EO-NoVM algorithm increases with the decrement of $T_{sw}$ as the loss of energy-efficiency present in EO-NoVM is directly proportional to $T_{sw}$ (\ref{ssec:moti}).       
\subsection{Performance evaluation}
The effects of delay constraint ($D_{max}^k$), scaling, round trip time ($T_{rtt}^k$) and the traffic shape parameter ($\alpha_{ON}$) on the performance (evaluated in term of energy-efficiency and average delay) of the proposed scheme is discussed in this subsection. For doing so, we choose same parameters as proposed in \ref{ssec:compa}.
\subsubsection{Effect of delay constraint}
The energy-efficiency and the average delay as a function of traffic load ($l$) for different values of delay constraints ($D_{max}^k (\forall k)=5ms,10ms,15ms $) are plotted in Fig. \ref{effect}(a). It can be observed from Fig. \ref{effect}(a) that with an increase in the delay constraint (relaxed delay bound), the achieved energy-efficiency increases. This is because, with increase of $D_{max}^k$, the time interval over which scheduling can be performed by satisfying the delay constraint also increases resulting in an increment of the opportunity of US data clubbing.  Moreover, we observe that the deviation of the energy-efficiency from the theoretical upper bound is more for both low and high traffic load scenarios. Whereas at medium traffic load ($\sim 0.6$), energy-efficiency nearly matches with its theoretical upper bound. The reason can be explained as follows. 

At low traffic loads, grant sizes are very small and therefore, the first bit of US data reaches the OLT after the latest scheduling horizon of all the wavelengths in most of the cases. This results in creation of scheduling voids. 
The grant size increases with an increase of $L$ (up to $L\sim 0.6$) resulting in an increment of the possibility of US data clubbing. Hence, the achieved energy-efficiency figures of EO-NoVM algorithm approach closer to the the theoretical upper bound with an increase in $L$. However, beyond a certain load, grant size become so large that the scheduling of ONUs by satisfying the delay constraint forces creation of extra voids resulting in an increment of the deviation of the achieved energy-efficiency figures from the theological upper bound. The minimum value of average delay at medium load (Fig. \ref{effect}(a)) further supports maximum US data clubbing and hence, minimum loss of energy-efficiency due to $T_{sw}$ since, in EO-NoVM, packets are scheduled before the latest time instant (limited by the delay constraint) only if there is any possibility of combaining the US data. 
\subsubsection{Effect of scaling}
The energy-efficiency and average delay as a function of $L$ for different values of $N$ and $W$ ($N=16$ \& $W=2,N=32$ \& $W=4,N=64 \& W=8$ while keeping the $N:W$ ratio same) are plotted in Fig. \ref{effect}(b). As the theoretical upper bound is decided by $N:W$ ratio (\ref{etamax}), it remains same for all the scenarios considered. It can be observed from Fig. \ref{effect}(b) that with an increase in $N$ (with $N:W$ fixed), the energy-efficiency improves since the increment of $N$, creates more opportunities of consolidating the US data. If the aggregated arrival rate to the OLT ($r_{agg}$) is same as $r_d$ or an integer multiple of $r_d$  ($pr_d$), then ideally $p$ active wavelengths should be sufficient for performing scheduling and it provides best US data clubbing within the active wavelengths. However, due to the round trip time and the delay constraint, voids are created. This results in a requirement of at least $p+1$ active wavelengths for scheduling the ONUs.  
The proposed EO-NoVM scheme provides best consolidation of US data at those values of $L$, for which $r_{agg}$ is similar to $r_d$ or $pr_d$, however, by using more than $p$ wavelengths. Thus, the energy-efficiency figures are closest to the theoretical upper bound (for $N=32$) at $L=0.32$, $0.64$ and $0.99$  when $r_agg\sim r_d,\text{ }2r_d$ or $3r_d$ respectively. Further, it can be observed from Fig. \ref{effect}(b) that the average delay decreases with an increase in $L$ up to a certain $L$. This is because, at low traffic load, deficiency of US data clubbing results in scheduling ONUs to the latest point by satisfying the delay constraint (\ref{ssec:EO-NoVM}). The increment of $L$ provides better data clubbing (\ref{ssec:compa}), and it causes reduction of average delay. With further increase in traffic load, the average delay increases. This happens due to the subsequent increase of average buffer occupancy as observed in a typical queuing discipline. Further, the delay vs. load follows a similar trend as observed in the difference of energy-efficiency with its theoretical bound.
\subsubsection{Effect of round-trip time}
The energy-efficiency and average delay as a function of $L$ for different value of $T_{rtt}^k$ $\forall k$ ($=0.2ms,0.5ms,2ms$) have been plotted in Fig. \ref{effect}(c).  The increment of $T_{rtt}^k$ causes increment in $TC_{min}^{k,j}$ $\forall j$. Since, scheduling in wavelength $j$ can only be performed in between $TC_{min}^{k,j}$ and $t_R^{k,q}+D_{q}^k$, the increment of $TC_{min}^{k,j}$ results in decrement of the duration over which scheduling can be perform. This leads to  reduction of the opportunity of US data clubbing and hence, degradation of energy-efficiency figures with the increment of $T_{rtt}^k$. This further causes increment of average delay with an increase in $T_{rtt}^k$ as well. 
\begin{figure}[h]
	\centering
	\includegraphics[scale=.755]{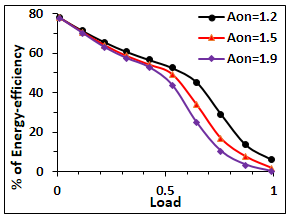}
	\caption{Effect traffic shape parameter ($\alpha_{ON}$) on energy-efficiency in the proposed EO-NoVM algorithm. Aon-$\alpha_{ON}$}
	\label{effect}
\end{figure} 
\subsubsection{Effect of traffic shape parameter ($\alpha_{ON}$)}
The energy-efficiency as a function of $L$ for different values of $\alpha_{ON}$ ($=1.2,1.5,1.9$) is plotted in Fig. 8. It can be observed from Fig. 8 that at a high load scenario, the energy-efficiency is significantly reduced for higher value of $\aleph_{ON}$. This is because, an increase in the value $\aleph_{ON}$ results in a subsequent increase in the variation of the size of the requested bandwidth. In such cases, the OLT scheduling creates more voids to satisfy the delay constraint.

\section{Conclusion}\label{conc}
In this paper, we have proposed an online TWDM based scheduling protocol for minimizing the energy-consumption of OLT receivers while satisfying up link delay constraint. In the process, we have demonstrated that minimization of the number of scheduling voids is more energy-efficient, compared to the conventional minimization of the number of active wavelengths \cite{online} (WM-\cite{online}). The improvement in energy-efficiency of the OLT receivers at high load is almost $25\%$ as compared to existing WM-\cite{online}. Further, we have provided an analytical upper bound of energy-efficiency for a given network load and demonstrated (using simulations) that the proposed EO-NoVM obtains an energy-efficiency figure close to the derived upper bound. Our simulation reveals that the energy-efficiency figures of EO-NoVM approach even closer to the theoretical upper bound with a decrease of sleep-to-wake-up time. We have further observed that a PON with more users and a smaller span is more energy-efficient.

\ifCLASSOPTIONcaptionsoff
  \newpage
\fi


\begin{thebibliography}{10}
	\providecommand{\url}[1]{#1}
	\csname url@rmstyle\endcsname
	\providecommand{\newblock}{\relax}
	\providecommand{\bibinfo}[2]{#2}
	\providecommand\BIBentrySTDinterwordspacing{\spaceskip=0pt\relax}
	\providecommand\BIBentryALTinterwordstretchfactor{4}
	\providecommand\BIBentryALTinterwordspacing{\spaceskip=\fontdimen2\font plus
		\BIBentryALTinterwordstretchfactor\fontdimen3\font minus
		\fontdimen4\font\relax}
	\providecommand\BIBforeignlanguage[2]{{%
			\expandafter\ifx\csname l@#1\endcsname\relax
			\typeout{** WARNING: IEEEtran.bst: No hyphenation pattern has been}%
			\typeout{** loaded for the language `#1'. Using the pattern for}%
			\typeout{** the default language instead.}%
			\else
			\language=\csname l@#1\endcsname
			\fi
			#2}}
	
	\bibitem{kani}
	J.-i. Kani, ``Power saving techniques and mechanisms for optical access
	networks systems,'' \emph{Journal of Lightwave Technology}, vol.~31, no.~4,
	pp. 563--570, 2013.
	
	\bibitem{bhar2015designing}
	C.~Bhar, N.~Chatur, A.~Mukhopadhyay, G.~Das, and D.~Datta, ``Designing a green
	optical network unit using arma-based traffic prediction,'' in \emph{Advanced
		Networks and Telecommuncations Systems (ANTS), 2015 IEEE International
		Conference on}.\hskip 1em plus 0.5em minus 0.4em\relax IEEE, 2015, pp. 1--6.
	
	\bibitem{doze}
	S.~Herreria-Alonso, M.~Rodriguez-Perez, M.~Fernandez-Veiga, and
	C.~Lopez-Garcia, ``On the use of the doze mode to reduce power consumption in
	epon systems,'' \emph{Journal of Lightwave Technology}, vol.~32, no.~2, pp.
	285--292, 2014.
	
	\bibitem{dixit}
	A.~Dixit, B.~Lannoo, D.~Colle, M.~Pickavet, and P.~Demeester, ``Energy
	efficient dynamic bandwidth allocation for ethernet passive optical networks:
	Overview, challenges, and solutions,'' \emph{Optical Switching and
		Networking}, vol.~18, pp. 169--179, 2015.
	
	\bibitem{Dixit:12}
	M.~P.~I. Dias, E.~Wong, D.~P. Van, and L.~Valcarenghi, ``Offline
	energy-efficient dynamic wavelength and bandwidth allocation algorithm for
	twdm-pons,'' in \emph{Communications (ICC), 2015 IEEE International
		Conference on}.\hskip 1em plus 0.5em minus 0.4em\relax IEEE, 2015, pp.
	5018--5023.
	
	\bibitem{dixit2013towards}
	A.~Dixit, S.~Lambert, B.~Lannoo, D.~Colle, M.~Pickavet, and P.~Demeester,
	``Towards energy efficiency in optical access networks,'' in \emph{Advanced
		Networks and Telecommuncations Systems (ANTS), 2013 IEEE International
		Conference on}.\hskip 1em plus 0.5em minus 0.4em\relax IEEE, 2013, pp. 1--6.
	
	\bibitem{khotimsky2014unifying}
	D.~A. Khotimsky, D.~Zhang, L.~Yuan, R.~O. Hirafuji, and D.~R. Campelo,
	``Unifying sleep and doze modes for energy-efficient pon systems,''
	\emph{IEEE Communications Letters}, vol.~18, no.~4, pp. 688--691, 2014.
	
	\bibitem{twdm}
	L.~Valcarenghi, Y.~Yoshida, A.~Maruta, P.~Castoldi, and K.-i. Kitayama,
	``Energy saving in twdm (a) pons: Challenges and opportunities,'' in
	\emph{Transparent Optical Networks (ICTON), 2013 15th International
		Conference on}.\hskip 1em plus 0.5em minus 0.4em\relax IEEE, 2013, pp. 1--4.
	
	\bibitem{online}
	P.~Garfias, S.~Sallent, L.~Guti{\'e}rrez, M.~De~Andrade, M.~Tornatore, and
	A.~Buttaboni, ``A novel traffic-aware mechanism for energy-saving at the olt
	in wdm/tdm-pon,'' in \emph{Network and Optical Communications (NOC), 2013
		18th European Conference on and Optical Cabling and Infrastructure (OC\&i),
		2013 8th Conference on}.\hskip 1em plus 0.5em minus 0.4em\relax IEEE, 2013,
	pp. 225--232.
	
	\bibitem{twdm1}
	W.~Xu, M.~Fu, and Z.~Le, ``Energy efficiency scheme for delay aware twdm-pon,''
	in \emph{Optical Communications and Networks (ICOCN), 2015 14th International
		Conference on}.\hskip 1em plus 0.5em minus 0.4em\relax IEEE, 2015, pp. 1--3.
	
	\bibitem{twdm2}
	M.~P.~I. Dias, D.~P. Van, L.~Valcarenghi, and E.~Wong, ``Energy-efficient
	dynamic wavelength and bandwidth allocation algorithm for twdm-pons with
	tunable vcsel onus,'' in \emph{Optical Fibre Technology, 2014 OptoElectronics
		and Communication Conference and Australian Conference on}.\hskip 1em plus
	0.5em minus 0.4em\relax IEEE, 2014, pp. 1007--1009.
	
	\bibitem{diajocn}
	------, ``Energy-efficient framework for time and wavelength division
	multiplexed passive optical networks,'' \emph{Journal of Optical
		Communications and Networking}, vol.~7, no.~6, pp. 496--504, 2015.
	
	\bibitem{verfy}
	M.~P.~I. Dias, E.~Wong, D.~P. Van, and L.~Valcarenghi, ``Offline
	energy-efficient dynamic wavelength and bandwidth allocation algorithm for
	twdm-pons,'' in \emph{Communications (ICC), 2015 IEEE International
		Conference on}.\hskip 1em plus 0.5em minus 0.4em\relax IEEE, 2015, pp.
	5018--5023.
	
	\bibitem{valcarenghi2015erratum}
	L.~Valcarenghi and P.~Castoldi, ``Erratum to icton 2014 paper tu. b4. 4: Twdm
	pon: How much energy can we really save?'' in \emph{Transparent Optical
		Networks (ICTON), 2015 17th International Conference on}.\hskip 1em plus
	0.5em minus 0.4em\relax IEEE, 2015, pp. 1--5.
	
	\bibitem{luo2012wavelength}
	Y.~Luo, M.~Sui, and F.~Effenberger, ``Wavelength management in time and
	wavelength division multiplexed passive optical networks (twdm-pons),'' in
	\emph{Global Communications Conference (GLOBECOM), 2012 IEEE}.\hskip 1em plus
	0.5em minus 0.4em\relax IEEE, 2012, pp. 2971--2976.
	
	\bibitem{eft}
	K.~Kanonakis and I.~Tomkos, ``Improving the efficiency of online upstream
	scheduling and wavelength assignment in hybrid wdm/tdma epon networks,''
	\emph{IEEE Journal on selected areas in Communications}, vol.~28, no.~6,
	2010.
	
	\bibitem{del}
	G.~Kramer, B.~Mukherjee, S.~Dixit, Y.~Ye, and R.~Hirth, ``Supporting
	differentiated classes of service in ethernet passive optical networks,''
	\emph{Journal of Optical Networking}, vol.~1, no.~8, pp. 280--298, 2002.
	
	\bibitem{Data}
	D.~P. Bertsekas and R.~G. Gallager, \emph{Data Networks}.\hskip 1em plus 0.5em
	minus 0.4em\relax Prentice Hall, 1992, ch.~3, p. 169.
	
	\bibitem{ieee2010ieee}
	I.~.~W. Group \emph{et~al.}, ``Ieee standard for information
	technology-telecommunications and information exchange between systems-local
	and metropolitan area networks-specific requirements part 11: Wireless lan
	medium access control (mac) and physical layer (phy) specifications,''
	\emph{IEEE Std}, vol. 802, no.~11, 2010.
	
	\bibitem{surve1}
	M.~P. McGarry, M.~Reisslein, and M.~Maier, ``Ethernet passive optical network
	architectures and dynamic bandwidth allocation algorithms,'' \emph{IEEE
		Communications Surveys \& Tutorials}, vol.~10, no.~3, 2008.
	
	\bibitem{mcgarry2004ethernet}
	M.~P. McGarry, M.~Maier, and M.~Reisslein, ``Ethernet pons: a survey of dynamic
	bandwidth allocation (dba) algorithms,'' \emph{IEEE communications magazine},
	vol.~42, no.~8, pp. S8--15, 2004.
	
	\bibitem{self}
	W.~E. Leland, M.~S. Taqqu, W.~Willinger, and D.~V. Wilson, ``On the
	self-similar nature of ethernet traffic (extended version),'' \emph{IEEE/ACM
		Transactions on networking}, vol.~2, no.~1, pp. 1--15, 1994.
	
\end{thebibliography}
\end{document}